# Astrochemical Bistability: Autocatalysis in Oxygen Chemistry

Gwénaëlle Dufour[1,2,3] and Steven B. Charnley[1]
[1] Astrochemistry Laboratory, Code 691, NASA Goddard Space Flight Center, Greenbelt, MD 20771, USA
[2] Department of Physics, Catholic University of America, Washington, DC 20064, USA
[3] Leiden Observatory, Leiden University, Leiden, The Netherlands

## Abstract

The origin of bistable solutions in the kinetic equations describing the chemistry of dense interstellar clouds is explained as being due to the autocatalysis and feedback of oxygen nuclei from the oxygen dimer ($O_2$). We identify four autocatalytic processes that can operate in dense molecular clouds, driven respectively by reactions of $H^+$, $He^+$, $C^+$, and $S^+$ with $O_2$. We show that these processes can produce the bistable solutions found in previous studies, as well as the dependence on various model parameters such as the helium ionization rate, the sulfur depletion and the $H_3^+$ electron recombination rate. We also show that ion–grain neutralizations are unlikely to affect the occurrence of bistability in dense clouds. It is pointed out that many chemical models of astronomical sources should have the potential to show bistable solutions.

## 1. Introduction

The occurrence of bistable solutions in models of dense cloud chemistry has been known for many years (Pineau des Forêts et al. 1992; Le Bourlot et al. 1993). These are characterized by the coexistence of two stable states connected by an unstable branch. The stable states occur in a molecular hydrogen gas and comprise a high ionization phase (the so-called HIP) in which the saturated species are under-abundant and protonation reactions are less efficient; and a low ionization phase in which saturated molecular ions are more abundant (the LIP). Several studies have shown that the occurrence of the bistable region is controlled by various combinations of the cosmic-ray ionization rates of $H_2$ and He, the gas density, the relative depletion of the heavy elements, interaction with dust grains, and the electron dissociative recombination rate of $H_3^+$ (Le Bourlot et al. 1993, 1995a, 1995b; Shalabiea & Greenberg 1995; Lee et al. 1998; Forêts & Roueff 2000; Viti et al. 2001; Charnley & Markwick 2003; Boger & Sternberg 2006; Wakelam et al. 2006). However, a definitive understanding of the origin of these bistable solutions has not emerged.

Bistability is of interest because it occurs within the region of parameter space that includes the depletions and densities relevant to dense interstellar clouds. It has been proposed as the origin of the high CI/CO ratios and low $O_2$ abundances observed in molecular clouds (Le Bourlot et al. 1993; Viti et al. 2001), in cloud chemistry near supernova remnants (Ceccarelli et al. 2011), in Galactic spiral arm clouds (Tieftrunk et al. 1994), and in high-latitude clouds (Gerin et al. 1997). Understanding the underlying bistable solutions may provide important insight into other astronomical environments.

An early explanation of the underlying chemical mechanism concerned an ionization instability connected to the difference in the respective rates of radiative and dissociative electron recombination of $H^+$ and $H_3^+$ (Pineau des Forêts et al. 1992). Boger & Sternberg (2006) also considered an ionization instability origin in which bistability could develop in an elementary $H_3^+$–$O_2$–$S^+$ cycle. Bistability cannot be solely due to the presence of $S^+$ since calculations show that bistable solutions occur even when S chemistry is neglected (Lee et al. 1998). It is well known in chemical kinetics that nonlinearity and feedback (i.e., autocatalysis) in the governing differential equations can lead to regions of multistability and complex chemical evolution (e.g., Epstein & Showalter 1996). Le Bourlot et al. (1993) suggested that autocatalysis could be responsible and here we demonstrate the fundamentally autocatalytic nature of astrochemical bistability.

## 2. Chemical Bistability in Dense Clouds

For prescribed initial conditions, the molecular evolution of a static dark cloud is obtained by solving a nonlinear system of ordinary differential equations for the abundances of $N$ chemicals, subject to conservation of charge and elemental nuclei (e.g., Nejad 2005). Bifurcations in the solutions are found by solving for the steady-state abundance

$$\boldsymbol{F}(\boldsymbol{x}; \zeta, n_H, T, \beta, \delta_j, \delta_M) = 0, \quad (1)$$

where $\zeta$ is the cosmic-ray ionization rate, $n_H$ is the number density of hydrogen nuclei ($\approx 2n(H_2)$), $T$ is the gas kinetic temperature, $\beta$ is the photorate associated with the cosmic-ray induced radiation field (Prasad & Tarafdar 1983; Gredel et al. 1989), $\delta_j$ are the depletion factors for the major volatile elements and $\delta_M$ is a depletion factor for refractory metals (Na in this paper). The fractional abundance of any species is $x(i) = n(i)/n_H$ and the total fractional abundance of each element, $j$, is given by $\delta_j X_j$, with $X_j$ relative to total H nuclei. An additional parameter known to influence bistability is the $H_3^+$ electron recombination rate, $\alpha_3$ (Pineau de Forêts & Roueff 2000). We initially adopted a value ten times larger than Millar et al. (1997) (Table 1), twice the value of $1.5 \times 10^{-7} T_3^{-0.5}$ cm$^3$ s$^{-1}$ used by Le Bourlot et al. and at the higher end of the range covered in previous studies ($T_3 = T/300$ K).

We solve the system of nonlinear Equations (1) by Newton–Raphson iteration and explore how the solutions undergo bifurcations as the control parameters are varied.

### 2.1. Model Calculations

To determine the underlying cause of bistability we begin with a known bistable solution and sequentially relax the $\delta_j$

**Table 1**
Dense Cloud Chemistry Model Parameters

| | | |
|---|---|---|
| Temperature | $T$ | 10 K |
| Visual extinction | $A_V$ | 15 mag |
| Cosmic-ray ionization rate | $\zeta$ | $5 \times 10^{-17}$ s$^{-1}$ |
| $H_3^+$ electron recombination rate[†] | $\alpha_3$ | $3 \times 10^{-7} T_3^{-0.5}$ cm$^3$ s$^{-1}$ |
| Elemental abundances[‡] | $X_O$ | $8.53 \times 10^{-4}$ |
| | $X_C$ | $3.62 \times 10^{-4}$ |
| | $X_S$ | $1.85 \times 10^{-5}$ |
| | $X_{Na}$ | $1.5 \times 10^{-8}$ |
| | $X_{He}$ | 0.1 |

**Table 2**
Bifurcation Models

| Model | $\delta_O$ | $\delta_C$ | $\delta_S$ | $\delta_{Na}$ | $\beta \neq 0$ | Note |
|---|---|---|---|---|---|---|
| 1 | 0.1 | 0.1 | 0.1 | 1.0 | Yes | Reference model |
| 2 | 0.1 | 0.1 | 0.01 | 1.0 | Yes | |
| 3 | 0.1 | 0.1 | 0 | 0 | Yes | |
| 4 | 0.1 | 0.1 | 0 | 0 | No | |
| 5 | 0.1 | 0 | 0 | 0 | No | |
| 6 | 0.1 | 0 | 0 | 0 | No | As Model 5 but $\delta_{He} = 0$ |

control parameter. Six models are sufficient to illustrate the processes leading to astrochemical bistability. Table 1 lists the physical conditions common to each model. The relevant parameters of each model are listed in Table 2. The chemical network is derived from Millar et al. (1997). Figures 1(a)–(f) show the fractional abundances of major chemical species as a function of $n_H$ in each of Models 1 to 6.

Model 1 is taken as a reference model (see Charnley & Markwick 2003 and Le Bourlot et al. 1993). It includes the internal radiation field with carbon, oxygen, and sulfur, depleted by a factor of 0.1. Figure 1(a) shows that the HIP/LIP bistable region occurs between $n_H \approx (1.8$–$4.9) \times 10^4$ cm$^{-3}$. The bifurcations consist of two stable solutions connected by an unstable one (e.g., Drazin 1992). In Model 2 the sulphur depletion is increased by a factor of 10, $\delta_S = 0.01$. Model 3 shows the effect of removing both S and Na from the network, $\delta_S = \delta_{Na} = 0$. In Model 4 the radiation field is turned off. Model 5 has carbon removed, $\delta_C = 0$, so that only chemical reactions involving oxygen, hydrogen, and helium remain. Finally, Model 6 is the same as Model 5 except that it contains no helium.

Figures 1(e) and (f) (Models 5 and 6) demonstrate that the cause of astrochemical bistability lies in the oxygen chemistry, and is unconnected to the presence of metals, of sulphur, or even of carbon.

### 2.2. Oxygen Chemistry: Autocatalysis and Feedback

We identify the fundamental mechanism that controls the occurrence of bistable solutions as autocatalysis (e.g., Gray & Scott 1994) in the reduced Models 5 and 6. The first step is formation of the oxygen dimer

$$O + OH \longrightarrow O_2 + H \quad (2)$$

followed by release of two oxygen nuclei in either of two *autocatalytic steps*

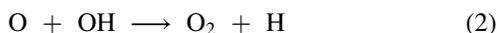
$$H^+ + O_2 \longrightarrow O_2^+ + H \quad (3)$$

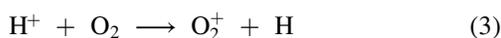
$$O_2^+ + e^- \longrightarrow O + O \quad (4)$$

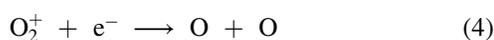

and

$$He^+ + O_2 \longrightarrow O^+ + O + He. \quad (5)$$

The oxygen ions and atoms then react to produce $OH^+$ to initiate two *feedback pathways*

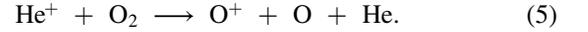
$$O^+ + H_2 \longrightarrow OH^+ + H \quad (6)$$

$$O + H_3^+ \longrightarrow OH^+ + H_2 \quad (7)$$

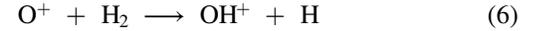
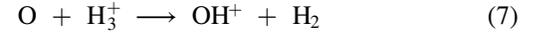

and subsequently become reincorporated into OH in the rapid sequence

$$OH^+ \xrightarrow{H_2} H_2O^+ \xrightarrow{H_2} H_3O^+ \xrightarrow{e^-} OH \quad (8)$$

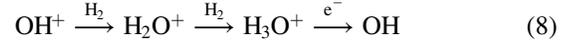

to complete the cycle. This is an example of *indirect autocatalysis* (e.g., Plasson et al. 2011).

Thus, we can identify two feedback routes to bistability: (1) an $O^+$ pathway initiated by $He^+$, and (2) a pure O atom pathway initiated by $H^+$. In Model 6 (no He) the O atom pathway is clearly solely responsible. As one moves across the bistable region from the HIP to the LIP in Model 5, Figure 1(e) shows that, although $H^+$ is the dominant ion in the HIP, close to the the bifurcation point in the LIP, $He^+$ is the dominant ion. Thus, $He^+$ plays an important role in producing bifurcations. Both pathways operate together in Model 5 but we find that one can come to dominate the other, depending on the value of $\alpha_3$.

Figure 2(a) shows that when the rate of reaction (5) is artificially set to zero in Model 5 the removal of autocatalysis and the $O^+$ pathway causes bistability to disappear. By contrast, in Model 6 (with no He), $n_e \approx n(H^+)$, autocatalysis is only through reactions (3) and (4) and the higher $H^+$ abundance drives reaction (3) more efficiently. As $n(H_3^+) \propto (\alpha_3 n_e)^{-1}$, bistability via the $O + H_3^+$ pathway can be recovered in the model of Figure 2(a) by reducing $\alpha_3$(10 K) from $1.64 \times 10^{-6}$ cm$^3$ s$^{-1}$ (see Table 1). Figure 2(b) shows the effect of reducing $\alpha_3$(10 K) by a factor of 10; the resulting higher $H_3^+$ abundance makes the O atom pathway more efficient and the bistable solution appears. Hence, in Model 5 whichever pathway produces bistability depends on the value of $\alpha_3$ relative to some critical value, $\alpha_{crit}$; by numerical experiment we find that $\alpha_{crit} \sim 10^{-6}$ cm$^3$ s$^{-1}$. Artificially setting the rate coefficient of reaction (4) to zero in either Model 5 or 6, removes the O atom pathway, causing bistability to disappear.

### 2.3. Dense Cloud Chemistry Reconstructed

In more realistic chemical models of molecular clouds, the inclusion of carbon and sulfur means that $C^+$ and $S^+$ can replace or complement $He^+$ and $H^+$ as the major ion destroying $O_2$. Figures 1(b)–(d) show that $C^+$ is the most abundant ion at the HIP and LIP. When carbon is added to the oxygen chemistry of Model 5, $C^+$ is formed by

$$He^+ + C \longrightarrow C^+ + He \quad (9)$$

and by reaction with CO. In Model 4 the autocatalytic step of reaction (5) is replaced by the net effect of the two product branchings in

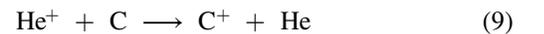
$$C^+ + O_2 \longrightarrow CO^+ + O \quad (10)$$

$$\longrightarrow CO + O^+. \quad (11)$$

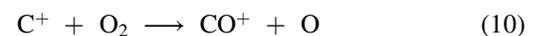
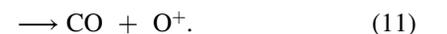

Although reaction (10) followed by

$$CO^+ + e^- \longrightarrow C + O \quad (12)$$

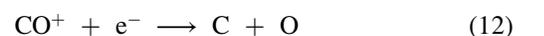

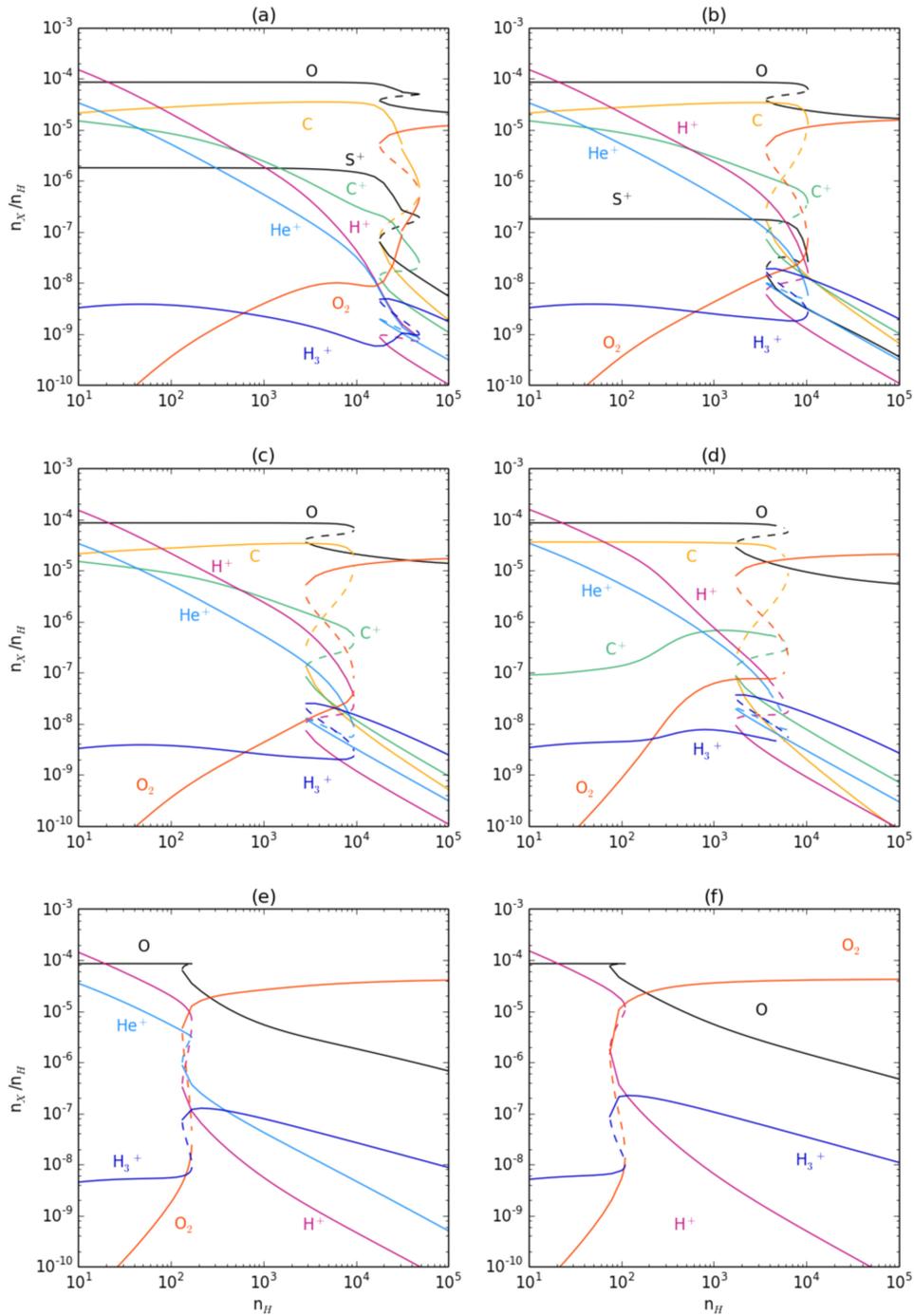

**Figure 1.** Computed steady-state abundances as a function of hydrogen density for Models 1–6: (a)–(f).

could in principle be an autocatalytic process, we find that the rapid destruction of $CO^+$ by $H_2$ renders its contribution negligible.

The autocatalytic nature of bistable solutions can also be confirmed through artificial models, as in Section 2.2. This is illustrated in Figure 2(c) which shows that the bifurcations in Model 4 (see Figure 1(a)) disappear once the rate coefficients of reactions, (10) and (11) have been set to zero. As in the case of the artificial models of Figures 2(a) and (b), bistability can be recovered by reducing the value of $\alpha_3$, as shown in Figure 2(d). In this case we find that reaction (5) is responsible for the bistable solution. We find that acting together these three reactions, (5), (10), and (11), are solely responsible for bistability in Model 4 independent of $\alpha_3$. In similarly artificial versions of Model 3 (not shown), we find the same autocatalytic effects except that, at low $\alpha_3$, $O_2^+$ ions from $O_2$ photoionization act in concert with reaction (4). In this model, at low $\alpha_3$, instead of reaction (5), reactions (3) and (4) combined with the $C^+ + O_2$ autocatalytic step, are responsible for the bistable solution.

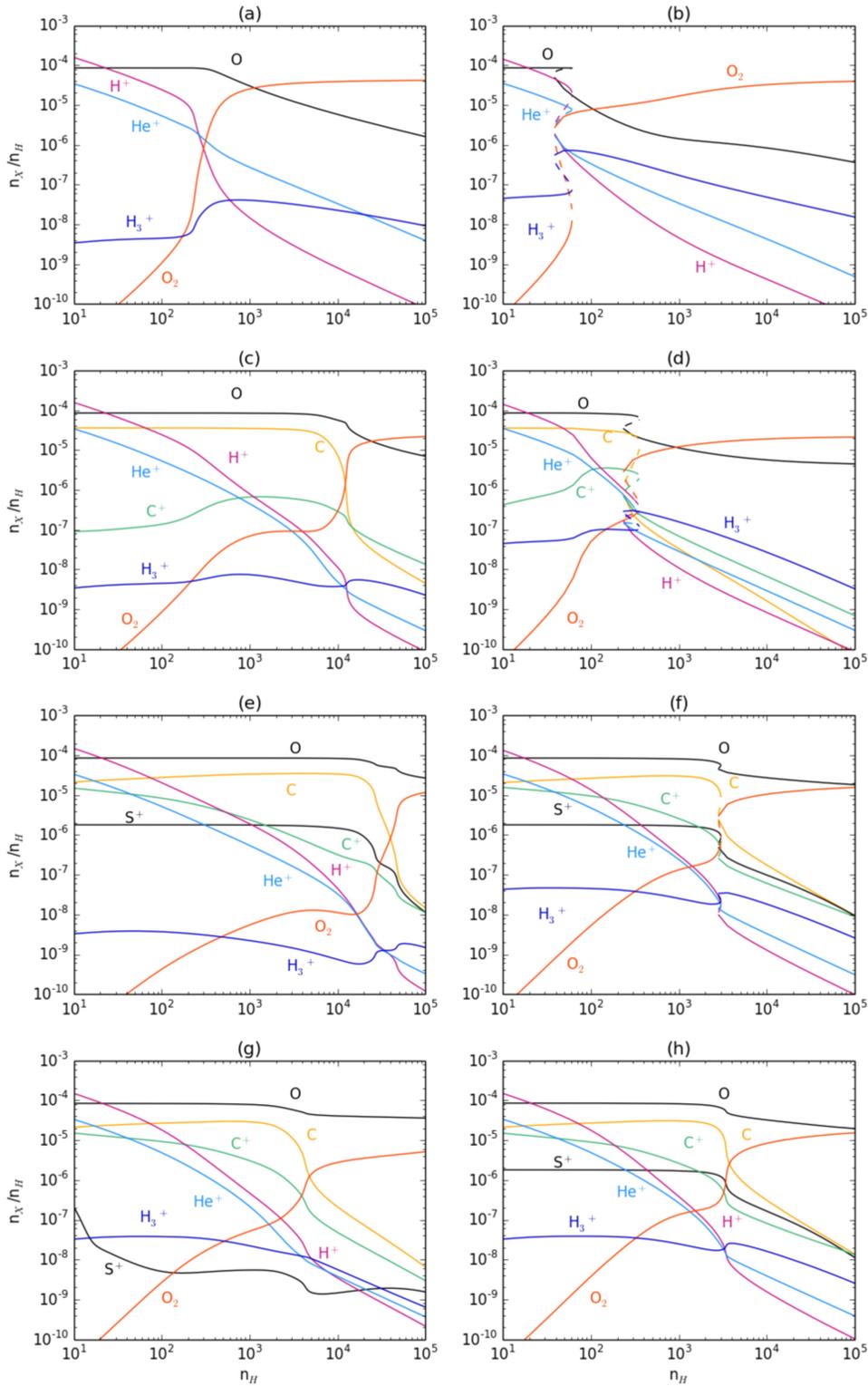

**Figure 2.** (a) Model 5 (Figure 1(e)) but with reaction (5) removed; (b) same as (a) except that $\alpha_3$ is reduced to $1.64 \times 10^{-7}$ cm$^3$ s$^{-1}$; (c) Model 4 (Figure 1(d)) but with reactions (10) and (11) removed. (d) Model 4 but with $\alpha_3$ reduced to $1.64 \times 10^{-7}$ cm$^3$ s$^{-1}$; (e) Model 1 (Figure 1(a)) but with reactions (10) and (11) removed; (f) same as (e) except that $\alpha_3$ is reduced to $1.64 \times 10^{-7}$ cm$^3$ s$^{-1}$; (g) same as (f) but with reaction (5) removed; (h) Model 1 (Figure 1(a)) but with only reaction (16) removed.

When sulphur is included at sufficiently high abundance, the reactions

$$C^+ + S \longrightarrow S^+ + C \quad (13)$$

$$H^+ + S \longrightarrow S^+ + H \quad (14)$$

can lead to $S^+$ being the dominant atomic ion in the bistable regime as shown in Model 1 (see Figure 1(a)), and the autocatalytic production of oxygen atoms can occur through

$$S^+ + O_2 \longrightarrow SO^+ + O \quad (15)$$

$$SO^+ + e^- \longrightarrow S + O. \quad (16)$$

In Model 2 the $\delta_S$ is sufficiently large that $S^+$ at the LIP bifurcation is not efficiently produced at the expense of $C^+$ (see Figure 1(b)) and the origin of the bistable solutions are the same as in Model 3.

Returning to the reference model (Model 1), Figure 2(e) shows that, as before, artificially removing only reactions (10) and (11) causes bistability to disappear in the high $\alpha_3$ case; Figure 2(f) shows that it again reappears at the lower value of $\alpha_3$. For Model 1 we have found that all the four identified autocatalytic processes can combine to act in concert in realistic models of molecular clouds. With the $C^+ + O_2$ autocatalytic reactions removed, as in Figure 2(f), further removing reaction (5) removes the bifurcation, as shown in Figure 2(g); this bifurcation also disappears if reaction (4) is removed instead of (5). Model 1 has the lowest S depletion and so $S^+$ can play an important rôle in bistability. Figure 2(h) shows the effect of only removing reaction (16).

Thus, the occurrence of bistability sensitively depends on four autocatalytic processes acting in unison: driven by $O_2$ reacting with $He^+$ and $C^+$, and with $H^+$ and $S^+$.

## 3. Discussion: Ion–grain Neutralization

These bistable solutions are produced in a gas-phase chemistry driven by cosmic rays. In dense molecular clouds most of the grains are negatively charged ($G^-$) and only about 4% are neutral ($G^0$) (e.g., Umebayashi & Nakano 1990). Atomic ion ($X^+$)-grain recombination

$$X^+ + G^- \longrightarrow X + G^0 \quad (17)$$

will dominate electron radiative recombination, and so could affect the onset of bistable solutions. For molecular ions, ion–grain recombination is never competitive with electron dissociative recombination in our models. Published studies that have considered the gas–grain interaction disagree as to whether or not bistable solutions can persist (Le Bourlot et al. 1995b; Wakelam et al. 2006), or are suppressed (Shalabiea & Greenberg 1995; Boger & Sternberg 2006). Shalabiea & Greenberg (1995) found that while the gas–grain interaction did suppress bistability for a model adopting diffuse ISM elemental depletions, bistability still occurred for the dense cloud values of Le Bourlot et al. (1993). Le Bourlot et al. (1995b) presented a more detailed treatment of the effects of the gas–grain interaction and confirmed that it had no effect in suppressing bistability. Boger & Sternberg (2006) also adopted diffuse ISM depletions and demonstrated that atomic ion–grain neutralization could indeed suppress their bistable solutions, whereas Wakelam et al. (2006) reported no effect.

We can assess the effect of ion–grain neutralization processes on our results as follows. The steady-state number density of an atomic ion, $X^+$, is

$$n(X^+) = \frac{P(X^+)}{L_{in} + k_i n(O_2) + \alpha_r n_e + \lambda(X^+)} \quad (18)$$

where $P(X^+)$ is the total ion production rate, $k_i$ and $\alpha_r$ are the rate coefficients for reactions with $O_2$ and electron radiative recombination, $L_{in}$ is the loss rate in all other ion–neutral reactions, and the ion–grain collision rate is $\lambda(X^+)$. Assuming a Maxwellian distribution for the ion velocities, and that all the grains are negatively charged (Umebayashi & Nakano 1990; Charnley 1997), we can write

$$\lambda(X^+) = \left(\frac{8kT}{\pi M_X m_H}\right)^{1/2} \langle n_{gr}(a)\sigma_c(a)\rangle, \quad (19)$$

where $k$ is the Boltzmann constant, $m_H$ is the proton mass, $T$ is the gas temperature (10 K), $M_X$ is the ion mass (in a.m.u), $n_{gr}(a)$ is the number density of grains of radius $a$, and $\sigma_c(a)$ is the collision cross-section including the Coulomb factor. We can write

$$\lambda(X^+) = C_{gr} M_X^{-1/2} n_H, \quad (20)$$

where $C_{gr}$ is the rate coefficient (cm$^3$ s$^{-1}$) for ion–grain collisions and is given by

$$C_{gr} = 1.45 \times 10^4 T^{1/2} \frac{\langle n_{gr}(a)\sigma_c(a)\rangle}{n_H}. \quad (21)$$

Grain neutralization can suppress a bistable solution produced by $X^+$ reacting with $O_2$ when

$$\lambda(X^+) > k_i n(O_2) \quad (22)$$

or when $C_{gr}$ exceeds a threshold value

$$C_{gr} > k_i M_X^{1/2} y(O_2), \quad (23)$$

where $y(O_2) = n(O_2)/n_H$ is the $O_2$ fractional abundance at the LIP bifurcation.

The bistable solutions are produced by $S^+$ in both the Shalabiea & Greenberg model (S/H = $9 \times 10^{-7}$) and in that of Boger & Sternberg (S/H $\approx 8 \times 10^{-6}$). When $X^+$ is $S^+$, as in Model 1, $k_i = 1.5 \times 10^{-11}$ cm$^3$ s$^{-1}$, $y(O_2) \approx 8 \times 10^{-6}$, and bistability is suppressed when $C_{gr} > 6.8 \times 10^{-16}$ cm$^3$ s$^{-1}$, in good agreement with the threshold calculated by Boger & Sternberg (2006).[4] However, in dense molecular clouds gaseous S is inferred to be depleted by a factor of $\sim$100 with respect to the diffuse ISM (e.g., Fuente et al. 2019), and so $C^+$ rather than $S^+$ drives bistability. When $X^+$ is $C^+$, as in Model 4, $k_i = 8.0 \times 10^{-10}$ cm$^3$ s$^{-1}$, $y(O_2) \approx 8 \times 10^{-6}$, and bistability is suppressed when $C_{gr} > 2.2 \times 10^{-14}$ cm$^3$ s$^{-1}$. Figure 3 shows the result of adding grain neutralization of atomic ions to Model 4, for two values of $C_{gr}$ that bracket the threshold value, and confirms that even a large value of $C_{gr}$ (5.6 $\times$ 10$^{-15}$ cm$^3$ s$^{-1}$) cannot suppress bistability.

How realistic are the $C_{gr}$ values required to cause bistable solutions to disappear? When $\langle n_{gr}(a) \sigma_c(a)\rangle$ corresponds to a single-size grain distribution with a dust/gas number density ratio appropriate for dense clouds (e.g., $n_{gr}(0.1\ \mu m) = 1.85 \times 10^{-12} n_H$) $C_{gr} \sim 3 \times 10^{-16}$ cm$^3$ s$^{-1}$; it is about seven times larger when an MRN grain size distribution (Mathis et al. 1977) is considered (Le Bourlot et al. 1995b). The latter value exceeds the $S^+$ threshold but neither value will have an effect on bistability arising from $C^+$. For the ion–grain collision rates[5] considered by Boger & Sternberg (2006), $C_{gr} \sim 2.5 \times 10^{-16}$–$2.5 \times 10^{-14}$ cm$^3$ s$^{-1}$, with bistability disappearing above $\approx 7.6 \times 10^{-16}\ \mathrm{cm^3\ s^{-1}}$, the largest collision rates ($\sim 10^{-14}\ \mathrm{cm^3\ s^{-1}}$) are unlikely to be relevant for dense clouds. These values pertain to MRN dust distributions that are appropriate to the diffuse interstellar medium, where the size distribution extends to very small carbonaceous particles, such as large PAH molecules (e.g., Weingartner & Draine 1999). Mid-IR observations of dense clouds however show clear evidence for very large dust grains ($\sim 1\ \mu$m) and no evidence for PAHs, suggesting that small grains have been removed from the distribution by coagulation (Steinacker et al. 2010). Thus, for realistic dense cloud elemental depletions and grain-size distributions, we conclude that the threshold for ion–grain neutralization to suppress bistability through $C^+$ reactions cannot be reached.

---
[4] Their rate coefficient for $S^+$-grain collisions at 50 K, $k_g$, is equal to $\sqrt{5/32} C_{gr}$ at 10 K.
[5] It is not possible to ascertain the value of $C_{gr}$ used in the model of Shalabiea & Greenberg.

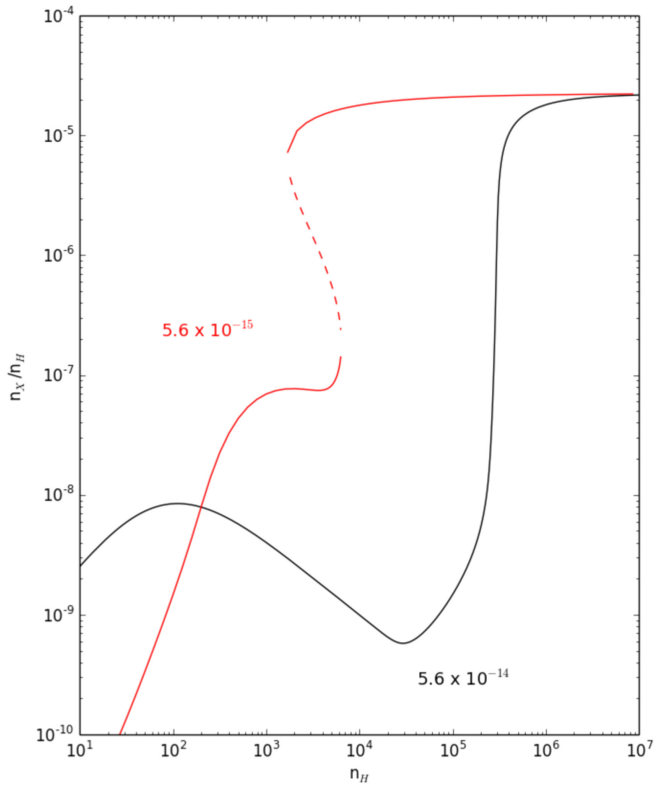

**Figure 3.** Molecular oxygen fractional abundance as a function of density in Model 4 for two different values of $C_{gr}$.

## 4. Conclusions

We have demonstrated that interstellar chemistry is bistable due to the interaction of several autocatalytic processes involving molecular oxygen. By deconstructing a known bistable solution into ever simpler reduced models through omission of chemical elements, and then artificially removing selected reactions, we have identified four distinct modes of autocatalysis that can occur in dense molecular clouds.

Our calculations provide explanations of results from earlier bistability studies: the dependence on C/O ratio (Le Bourlot et al. 1995a), the $He^+$ production rate (Wakelam et al. 2006), and the reasons for bistability appearing with or without sulfur (Lee et al. 1998). Apart from the previously known dependence of bistability on the $H_3^+$ recombination rate (Forêts & Roueff 2000), we find that $\alpha_3$ can also mediate between the feedback pathways (via $O^+$ or O) in the autocatalytic cycle.

Internally generated UV photons play a minor role in these models (although their effects do become more pronounced in artificial models, see Section 2.3). Refractory metals have no autocatalytic function and bistable solutions can occur in their absence.

The bistable solutions discovered in dark clouds are controlled by $\zeta/n_H$, the relative elemental depletions, and the value of the $H_3^+$ electron recombination rate. Recent measurements of the elemental ratios C/H and C/O in the dark cloud TMC-1 are consistent with those where bistability can occur (Fuente et al. 2019). The associated S/H ratio in TMC-1 is very low, as is commonly found in molecular clouds, indicating that autocatalysis involving S will not be as important as the competing processes (see Model 2). Photoprocesses have a relatively minor effect in these models but could become important at lower $A_V$.

We have shown that previous conclusions regarding the efficiency of atomic ion–grain neutralization in suppressing dense cloud bistability depend on the applicability of elemental depletions and dust distributions that are more appropriate for the diffuse ISM. For more realistic dense cloud model parameters (high S depletion and reduced number of small grains), grain recombination has no effect on bistable solutions.

The bistable solutions present in dense cloud chemistry are due to autocatalysis and not an ionization instability.[6] In fact, Le Bourlot et al. (1993, 1995a), in the first explicit demonstration of bistability, presciently suggested that oxygen autocatalysis through reactions (2)–(4) could play a role in interstellar bistability, as confirmed in the calculation presented here (Model 6, Figure 1(f)). Remarkably, this insight was neglected in all subsequent studies.

Gas-phase bistability is possible when an autocatalyst can form a dimer that is subsequently destroyed in an autocatalytic step followed by reformation of the dimer. As many such reactions are fundamental to chemical models of astronomical sources, bistability can be expected to be a common phenomenon (G. Dufour & S. B. Charnley 2019, in preparation).

This research was supported by NASA's Planetary Science Division Internal Scientist Funding Program through the Fundamental Laboratory Research work package (FLaRe).

---

[6] Boger & Sternberg (2006) attributed the origin of bistability as being due to an instability proposed by Pineau des Forêts et al. (1992). When compared to our results, we can see that their $H_3^+$–$O_2$–$S^+$ cycle is in fact autocatalytic.